\documentclass[12pt,a4paper]{article}
\usepackage{epsfig}
\def\beq{\begin{equation}}
\def\eeq{\end{equation}}
\def\bea{\begin{eqnarray}}
\def\eea{\end{eqnarray}}

\def\({\left(}   
\def\){\right)}   

\def\eq#1{{Eq.~(\ref{#1})}}

\def\arnps#1#2#3{  {\it Ann. Rev. Nucl. Part. Sci. }{\bf #1} (19#2) #3}
\def\npb#1#2#3{    {\it Nucl. Phys. }{\bf B#1} (19#2) #3}
\def\plb#1#2#3{    {\it Phys. Lett. }{\bf B#1} (19#2) #3}
\def\prd#1#2#3{    {\it Phys. Rev. }{\bf D#1} (19#2) #3}
\def\prep#1#2#3{   {\it Phys. Rep. }{\bf #1} (19#2) #3}
\def\prl#1#2#3{    {\it Phys. Rev. Lett. }{\bf #1} (19#2) #3}

\def\zpc#1#2#3{    {\it Z. Phys. }{\bf C#1} (19#2) #3}

\def\epj#1#2#3{    {\it Eur. Phys. J. }{\bf C#1} (19#2) #3}

\begin{document}

\title { {\LARGE\bf{SCREENING CORECTIONS IN   DIS}}\\[2ex]    
{\LARGE\bf{AT LOW $\mathbf{Q^2}$ and $\mathbf{x}$ }} }
\author{
{\bf
E.~Gotsman\thanks{E-mail: gotsman@post.tau.ac.il}~$\,^{a}$,
\quad E. ~Ferreira \thanks{E-mail: erasmo@if.ufrj.br}~$\,^{b}$,
\quad  E.~Levin\thanks{E-mail: leving@post.tau.ac.il,
 elevin@quark.phy.bnl.gov}~$\, ^{a}$,
}\\%[5mm]
{\bf 
 U.~Maor\thanks{E-mail: maor@post.tau.ac.il}~$\,^{a}$
 \quad and \quad
E. Naftali\thanks{E-mail: erann@post.tau.ac.}~$\,^{a}$ 
}\\[10mm]
 {\it\normalsize $^a$HEP Department}\\
 {\it\normalsize School of Physics and Astronomy,}\\ 
 {\it\normalsize Raymond and Beverly Sackler Faculty of Exact Science,}\\
 {\it\normalsize Tel-Aviv University, Ramat Aviv, 69978, Israel}\\[0.5cm]
{\it\normalsize $^b$  Instituto de Fisica, Universidade Federal do 
Rio de Janeiro}\\
{\it \normalsize Rio de Janeiro RJ21945-970, BRASIL}
}
\date{July 2000}
\maketitle
\thispagestyle{empty}
\begin{flushright}
\vspace{-15cm}
{\bf ICHEP2000\,\,,}\\
{\bf  paper No.489}\\
{\bf July 2000}
\end{flushright}

\newpage 
\begin{abstract}
We expect that s-channel unitarity should materialize
in hard DIS reactions through screening corrections (SC) indicating that
the gluon distribution function is approaching saturation,
it is not as yet clear what  the kinematical scales are at
which these effects become important. 
While the global DIS $\gamma^*p$ total cross section, or $F_2(x,Q^2)$,
data are well reproduced by DGLAP evolution without substantial SC,
 there exists   experimental data from HERA which suggests
deviations from DGLAP predictions in the small $Q^2$ and $x$ limits. These
signatures are observed in both the fine details of $F_2(x,Q^2)$ provided
$Q^2$ and $x$ are small enough, as well as in the diffractive
channels. In this investigation we present a detailed study of 
$\partial F_2/\partial \ln Q^2$ which is supported by a coupled analysis
of $J/\Psi$ photo and DIS production. Both channels are directly
proportional 
to $xG(x,Q^2)$, and as such serve as excellent discriminators
between different approaches and models. 
In the first phase of our
investigation we have found that none of the latest editions of the parton
distribution functions (GRV98, MRS99, CTEQ5) provides an adequate and
simultaneous reproduction of 
$Q^2$ logarithmic slope of $F_2$ at small $Q^2$ values as well as $J/\Psi$
photoproduction (Details of this will be published separately
\cite{GLMNF}). We then 
show that  taking GRV98NLO as input and correcting it for SC, we can
reproduce the recent HERA data well. The calculation depends on one
parameter $R^2=8.5GeV^{-2}$ which is directly determined from the $J/\Psi$
photoproduction forward differential slope. With this input we obtain an
excellent fit to the 
$J/\Psi$ photo and DIS production data. Our calculations made in the LLA
of pQCD take into account the corrections implied by the real part of the
production amplitude, off diagonal (skewed) gluon distributions and the
Fermi motion of the charm quarks within the bound Charmonium system. The
SC are consistently calculated for both the percolation of a $q\bar q$
through the target and the screening of the gluon parton distribution
which forms the base of our calculation.
Our main conclusion is that, whereas we find strong support for the need
for SC in the small $Q^2$ and $x$ limits of the channels we have
investigated,
the latest HERA data is not sufficiently precise   
to directly determine the gluon saturation scale.
\end{abstract}

\thispagestyle{empty}

\newpage
\setcounter{page}{1}

\section{Introduction}

Over the past few years we have been witness to vigorous experimental,
phenomenological and theoretical investigations of the proton 
deep inelastic scattering (DIS) structure functions and 
some exclusive channels with small $Q^2<5GeV^2$ and $x<10^{-2}$.
These comprehensive studies aim at establishing the applicability and
possible need for a re-formulation of pQCD, 
as we know it, when approaching the kinematic interface with the less
understood npQCD dominated domain.
The standard procedure for the pQCD analysis of DIS on a nucleon target 
has been to utilize
the DGLAP evolution equations for the structure functions as the key
ingredient for fixing the parton distribution functions (p.d.f.).
These p.d.f. are then used as input for the calculations of exclusive DIS
channels, usually executed in the color dipole approximation. 

The physics of small $Q^2$ and small $x$ is
associated with
the search for the scale of gluon saturation implied by s-channel unitarity
\cite{SAT}. 
One should remember, though, that gluon saturation signals the
transition from perturbative to non perturbative QCD. 
This transition is preceded   
by SC signatures which are expected to be
experimentally visible even though the relevant scattering amplitude has 
not yet reached the unitarity black limit. Moreover, from our
experience with
soft Pomeron physics, we know that   
different channels have different
scales at which  unitarity corrections become appreciable. 
 Specifically, the scale associated with the diffractive
channels are considerably smaller than those associated with the elastic
channel \cite{GLMprd}.  

%Considering the theoretical studies of DIS, it is
%not clear at 
%what  scales  unitarity effects, such as SC, become
%significant.

 Inspite of  significant theoretical progress in recent years
\cite{THEORY}, it is still not clear  what  the saturation scale
in the present 
experimentally accesible kinematic region is.
We recall that, while the global 
analysis of $F_2(x,Q^2)$ (or $\sigma_{tot}^{\gamma^*p}(W,Q^2)$) data 
shows no conclusive deviations from DGLAP, there are dedicated HERA 
investigations suggesting deviations from the DGLAP expectations in the
small $Q^2$ and $x$ limits. These signatures are observed in both the fine 
details of $F_2(x,Q^2)$ as well as in the diffractive channels, 
provided $Q^2$ and $x$ are small enough.

In the following we present a detailed study of 
$\partial F_2/\partial \ln Q^2$ which is supported by a coupled analysis
of $J/\Psi$ photo and DIS production. The strategy of our investigation is
based on the observation that these observables ( in LLA of pQCD )
 are linear in $xG(x,Q^2)$
and $\(xG(x,Q^2)\)^2$ respectively, and being relatively well measured 
may serve as effective discriminators when we compare their
detailed features with existing relevant theoretical approaches and
models. In the first phase of our investigation, we found
that none of the latest p.d.f. 
 \cite{GRV98}\cite{MRS99}\cite{CTEQ5} 
provides an adequate simultaneous reproduction, at small pQCD scales, 
of the recent HERA data on  
the logarithmic slope of $F_2$ \cite{H1slope}\cite{ZEUSslope} at small
$Q^2$ and $x$, 
as well as the abundant high energy data on $J/\Psi$ photo and DIS 
production \cite{ZEUSJ}\cite{H1J}\cite{fixedJ}. 
We then proceed to show that when GRV98NLO is corrected for
SC, as suggested in our previous 
publications \cite{GLMslope}\cite{GLMNslope}, it gives a good description
of
these data. The SC calculation depends on one parameter, 
$R^2=8.5GeV^{-2}$, which is directly deduced from the $J/\Psi$
photoproduction forward
differential slope. We elaborate on a recent suggestion \cite{G-BW} that 
the gluon saturation scale may be determined by examining the behaviour of 
$\partial F_2(x,Q^2)/\partial lnQ^2$ against $Q^2$ and $x$ at fixed W
values. In our opinion, this suggestion which was discussed also in a
recent presentation of the ZEUS new data \cite{ZEUSslope}, 
reflects  the particular kinematic
relationship between $Q^2, W^2$ and $x$ and does not provide an
unambiguous determination of the desired scale.   

Our approach can be tested in the high energy analysis of photo and DIS
exclusive
production of $J/\Psi$. This is seemingly a straight forward procedure as
the LLA pQCD calculation of this cross section is proportional to 
$\(xG(x,\bar Q^2)\)^2$, where $x$ and $\bar Q^2$ are determined from
$m_c$, the c-quark mass, and W, the c.m. energy. 
 A realistic calculation
depends on a few
corrections to the original pQCD estimate: $C_R^2$ due to the amplitude
real part, $C_g^2$ 
due to the contribution of the  
off diagonal (skewed) gluon distribution \cite{offdiagonal} 
and $C_F^2$ due to the Fermi motion 
deviations from the simple static non relativistic estimate of the
$J/\Psi$ wave function \cite{Fermi}. All of these corrections contribute
to the final
(amplitude squared) 
estimate. We present a detailed analysis of this channel which is
consistent with
 our approach and input choices made in the analysis of 
$\partial F_2/\partial \ln Q^2$.

We conclude with some general comments and observations on the 
outstanding problems of
gluon saturation and screening corrections.

\section{The small $Q^2$ and $x$ behavior of 
$\partial F_2/\partial \ln Q^2$}

Checking for unitarity corrections in $F_2$ studies is not simple. As is
well known, a global
DGLAP analysis of the data with the recent p.d.f.   
\cite{GRV98}\cite{MRS99}\cite{CTEQ5} is adequate. A study \cite{ADL}
comparing the screened and non screened DGLAP calculations of $F_2(x,Q^2)$ 
showed only a small difference due to SC even in the small $Q^2$ and $x$
attained by
present HERA measurements. Clearly, a unitarity study in the above
kinematic limit requires a dedicated investigation, confined to small
$Q^2$ and $x$, which can magnify the presumed experimental signatures.
We recall that in the small $x$ limit of DGLAP we have
\beq\label{A}
\frac{\partial F_2(x,Q^2)}{\partial ln Q^2}\,=
\,\frac{2\alpha_S}{9\pi}xG^{DGLAP}(x,Q^2).
\eeq
Accordingly, a significant deviation of the data from \eq{A},
where $xG^{DGLAP}$ is obtained from the global $F_2$ analysis,   
may serve as an experimental signature indicating the growing 
importance of unitarity corrections. 
This was first suggested by Caldwell \cite{caldwell}, showing a rather  
complicated plot of $\partial F_2/\partial \ln Q^2$ in which each point
had different $Q^2$ and $x$ values.
The Caldwell plot suggested a dramatic turn over of 
$\partial F_2/\partial ln Q^2$ corresponding to $Q^2$ of about $3 GeV^2$ 
and $x<5 \cdot 10^{-3}$ in contrast to the behavior expected 
from GRV94 \cite{GRV94} at sufficiently small $Q^2$ and $x$. The
problem with this presentation is that, as suggestive as it may seem, it
does not discriminate between different dynamical interpretations 
\cite{GLMslope}\cite{GLMNslope}\cite{G-BW}\cite{CKMT}\cite{DL2P}. 
It is actually compatible with an overall data generator \cite{ALLM} as
well as the latest p.d.f.  which were re-adjusted to
account for
this observation. We conclude  
that, where as a reproduction of the Cadwell plot is required as 
a pre condition for a serious consideration of any suggested model, the
plot on its own can not serve as an effective discriminator between models 
even in the extreme case when they are fundamentally different. 

A far better discrimination is obtained if we carefully
study the small $Q^2$ and $x$
dependences of $\partial F_2/\partial ln Q^2$ at either fixed $Q^2$ 
or fixed $x$ values being free from the kinematic correlation between
$Q^2$ and $x$ that plagued the Caldwell plot. Such preliminary HERA data 
have recently became available \cite{H1slope}\cite{ZEUSslope}. As we
shall show, a pQCD analysis of these data is consistent with   a
SC interpretation \cite{GLMNslope}.
 
We follow the eikonal SC formalism presented in Ref.\cite{GLMNslope}, 
where
screening is calculated in both the quark sector, to account for the
percolation of a $q\bar q$ through the target, and the gluon sector, to
account for the screening of $xG(x,Q^2)$. The factorizable result that we
obtain is
\beq\label{B}
\frac{\partial F_2^{SC}(x,Q^2)}{\partial ln Q^2}\,=
\,D_q(x,Q^2)D_g(x,Q^2)
\frac{\partial F_2^{DGLAP}(x,Q^2)}{\partial ln Q^2}.
\eeq

SC in the quark section are given by
\beq\label{C1}
D_q(x,Q^2)\frac{\partial F_2^{DGLAP}(x,Q^2)}{\partial ln Q^2}\,=
\,\frac{Q^2}{3\pi^2} \int\,db^2\(1\,-\,e^{-\kappa_q(x,Q^2;b^2)}\),
\eeq
\beq\label{C2}
\kappa_q\,=\,\frac{2\pi \alpha_S}{3Q^2}xG^{DGLAP}(x,Q^2)\Gamma(b^2).
\eeq
The calculation is significantly simplified if we assume a Gaussian 
parameterization for the two gluon non perturbative form factor,
\beq\label{C3}
\Gamma(b^2)\,=\,\frac{1}{R^2}e^{-b^2/R^2}.
\eeq

SC in the gluon sector are given by 
\beq\label{D1}
xG^{SC}(x,Q^2)\,=\,D_g(x,Q^2) xG^{DGLAP}(x,Q^2),
\eeq
where
\beq\label{D2}
xG^{SC}(x,Q^2)\,=\,
\frac{2}{\pi^2}\,\int_{x}^{1}\,\frac{dx^{\prime}}{x^{\prime}}
\int_{0}^{Q^2}\,dQ^{\prime\,2}\,
\int db^2\,\(1\,-\,e^{- \kappa_g(x^{\prime},Q^{\prime\,2};b^2)}\).
\eeq
Note that 
$\kappa_g(x^{\prime},Q^{\prime\,2};b^2)\,=\,
\frac{4}{9}\kappa_q(x^{\prime},Q^{\prime\,2};b^2)$ defined in \eq{C2}.
An obvious difficulty in the above calculation of $xG^{SC}$ stems from the
fact that the $Q^{\prime\,2}$ integration spans not only the short (pQCD),
but also the long (npQCD) distances. To overcome this difficulty we assume
that
\beq\label{D3}
xG\(x,Q^2\,<\,\mu^2\)\,=\,\frac{Q^2}{\mu^2}xG\(x,Q^2\),
\eeq 
where $\mu^2 \simeq Q_0^2$. Our choice of the above interpolation is
motivated by the gauge invariance requirement 
that $xG\,\propto\,Q^2$ when $Q^2\,\rightarrow\,0$.

The SC calculation described above can be applied to any given input 
p.d.f.
where the only adjusted parameters are $R^2$ and $\mu^2$. As we shall see
in the next section, $R^2 = 8.5 GeV^{-2}$ is determined directly from the
forward slope of $J/\Psi$ photoproduction. 
$\mu^2$ is conveniently fixed at $Q_0^2$, the lowest $Q^2$ value
accessible for  the input p.d.f. we use. Once we have chosen our p.d.f.,
our SC
calculation is essentially parameter free.  
We have checked that our output results are not sensitive to the
fine tuning of these fixed parameters. 

Our  results are presented
in Figs. 1,2 and 3. Throughout this investigation we have used as input
the $\bar {MS}$ version of GRV98NLO. Using the GRV98DIS version provides
very similar results. Following are some comments relating to our
calculations and results:
\newline 
1) As can be seen, in the limit of small $Q^2$ and $x$ there is a
significant difference between the screened and non screened values of 
$\partial F_2/\partial ln Q^2$. As expected the SC results are smaller and
softer than the non screened input. 
2) Visibly, our overall reproduction of the experimental data is very
good, in
particular when  considering that our input is essentially parameter free.
A proper $\chi^2$ calculation requires the knowledge of the
unknown theoretical errors.  
3) If we follow the standard procedure and replace
the theoretical errors with the experimental ones, we obtain excellent
$\chi^2/ndf=0.75$ for 21 H1 data points with the exception of 3
points which are visibly out of line. The ZEUS data has  errors which are
considerably smaller and as a result our $\chi^2/ndf$ is not as good, 
even though our reproduction of the ZEUS data is  reasonable.
\newline
4) The ZEUS $Q^2=1.9GeV^2$ and H1 $Q^2=3.0GeV^2$ 
data are somewhat softer than our
predictions, which do not contain  a soft non perturbative background.  

Golec-Biernat and Wuesthoff \cite{G-BW} have suggested studying the
$Q^2$ and $x$ behaviour of $\partial F_2/\partial ln Q^2$ at fixed W 
as a method to determine the gluon saturation scales from the anticipated
turn over in these plots.
Recent ZEUS low $Q^2$ presentations \cite{ZEUSslope} of these plots show,
indeed, the anticipated turn over structure in these figures, seemingly
suggesting that gluon saturation is attained at $Q^2\,\simeq \, 1 GeV^2$.
In our opinion the proper variables to study $F_2$ are $x$ and $Q^2$.
Trying to study the structure function by introducing other 
variables, such as W,
may result in spurious effects which are predominantly kinematic.
In the specific procedure suggested by Golec-Biernat and Wuesthoff, the
combination of the kinematic relation between $x, Q^2$ and W with the very
general behaviour of $xG(x,Q^2)$ is sufficient to produce a turn over. Its
exact location depends on the details of the numerical
input. Consequently, the
suggested fixed W plots do not convey any new dynamical information 
even if such information is hidden in the analyzed data. Actually, it
seems
that any $F_2$ (or $xG$) parameterization which factorizes the $Q^2$
and $x$ dependences, such as the Buchmueller- Haidt model \cite{H}, is
capable of
producing the fixed W turn over effects.

 To illustrate this point we consider the fixed W behaviour in two
models which have very different dynamics:
\newline
1)  Our own GLMN \cite{GLMNslope}, which is a pure pQCD dipole model
with
SC.
As such, our model relates indirectly to gluon saturation, even though it
is constructed so as to include unitarity corrections below actual
saturation.
\newline
2)  The DL two Pomeron parametrization
\cite{DL2P}, which is based on the Regge formalism, and consists of the
coherent sum of contributions from a "hard" and a "soft" Pomeron, a normal
Reggeon and an additional contribution from the charmed sector which is
proportional to the "hard" Pomeron. Each of these fixed j-poles are
multiplied by a fitted $Q^{2}$ form factor. The parameters of the "model"
were determined from the requirement of a "best fit" to the experimental
$F_{2}(x,Q^{2})$ data, and are not directly associated with the gluon
distribution in the proton or "saturation".

 The results obtained for the 
logarithmic slope of $F_{2}$ at fixed W for both parametrizations are
compared to the ZEUS experimental results in Fig.4 and 6.
In Figs.5 and 7 we also display the behaviour of 
 $\partial F_2/\partial lnQ^2$ at fixed $Q^{2}$ and fixed x.
 We note that
both "models" provide a reasonable description of the ZEUS data including
the observed turn over in the fixed W plots.As GRV98 is only applicable
for $Q_0^2 \geq 1GeV^2$, we have repeated the
GLMN calculation with GRV94 which has $Q_0^2=0.4GeV^2$ and again
reproduce
the ZEUS fixed W turn over effect.
 \newline
From the above it appears that any model which provides a reasonable
description of $F_{2}(x,Q^{2})$ will exhibit a turn over in fixed W plots
of the logarithmic derivative of  $F_{2}(x,Q^{2})$, this occurs due to
the relation between the kinematic variables x, $Q^{2}$ and W, and is not
related to "saturation".

 However, the examples (the Buchmueller-Haidt model and
Donnachie-Landshoff approach)  which have been used to demonstrate that
the ``saturation" is not a unique mechanism for a turn over at fixed $W$,
have one common feature: the ``soft" contributions are concentrated at
a  rather large typical scale $\geq \,\,2 \,GeV^2$. This observation 
 supports  the idea that the so called soft Pomeron stems from
rather short distances \cite{KL}.

Therefore, an equally good description of the experimental data  can
be    achieved  either from gluon ``saturation" ( shadowing corrections
) or by 
matching soft and hard processes if the soft ones occur at rather short
distances $\leq\,\,0.5 \,GeV^{-2}$.

In conclusion, whereas we find strong support for the need for
SC in the small $Q^2$ and $x$ limits for $\partial F_2/\partial lnQ^2$
 we are unable, as yet, to
determine the gluon saturation scale directly from the latest HERA data.
The gluon saturation scale may be theoretically estimated from the
contours produced at the
boundary of $\kappa_g\,=\,1$, as discussed in our 
papers \cite{GLMslope}\cite{GLMNslope}.

\begin{figure} \centerline{\epsfig{file=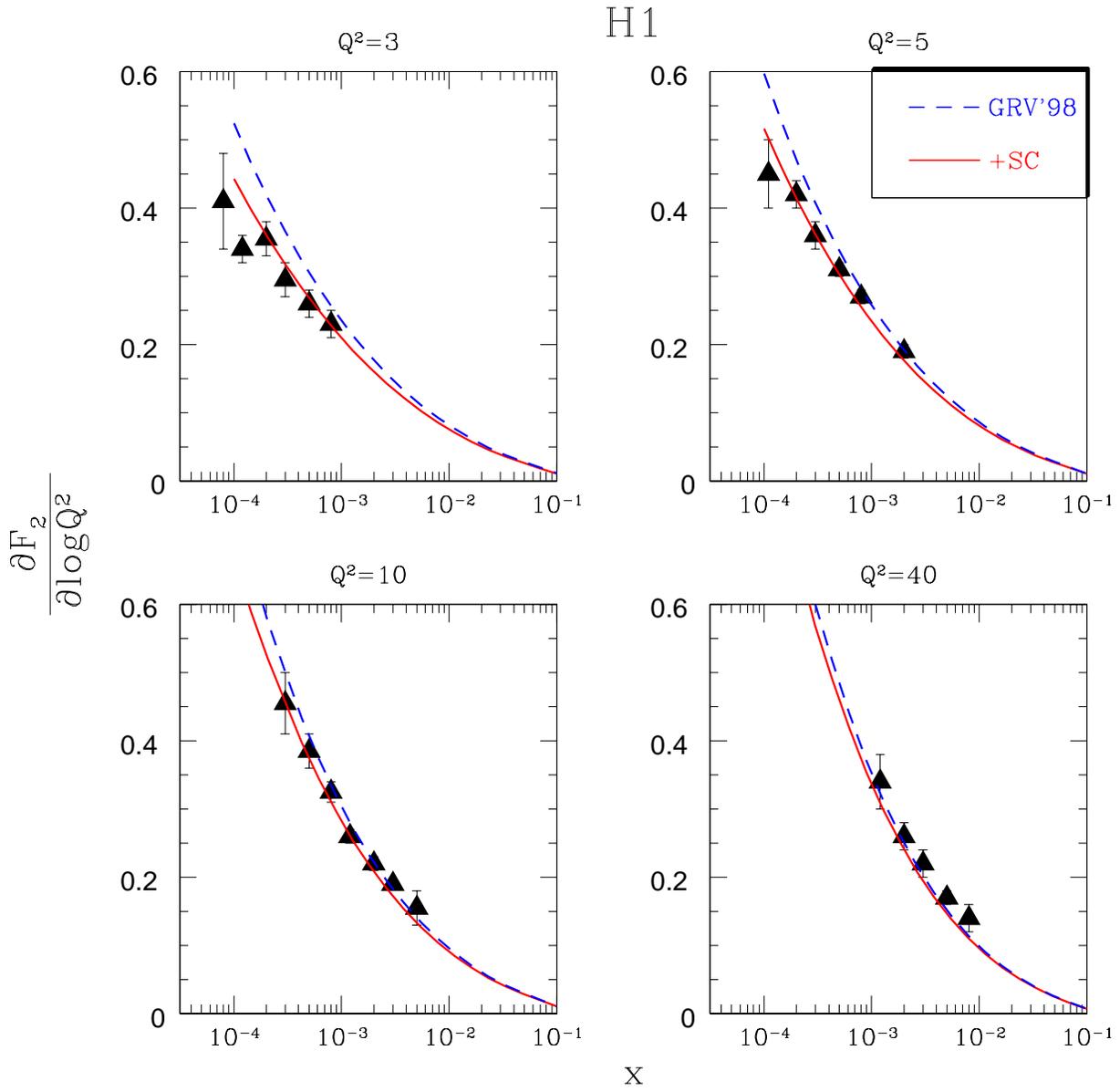,width=170mm}}
\caption{\it x dependence of H1 logarithmic slope data at fixed $Q^2$
compared with our calculations.}
\label{Fig.1}
\end{figure}

\begin{figure}
\centerline{\epsfig{file=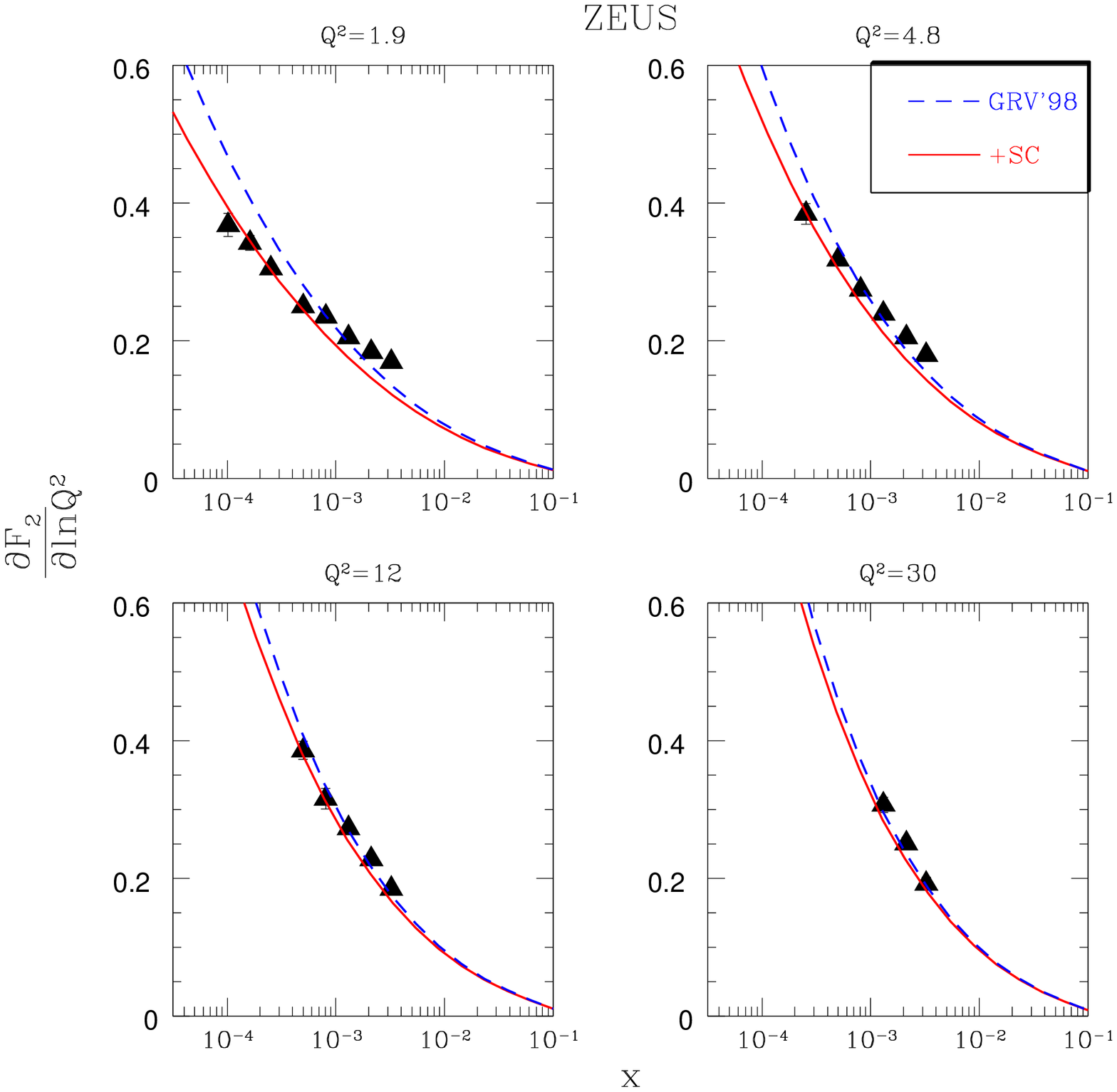,width=170mm}}
\caption{\it  x dependence of ZEUS logarithmic slope data at fixed $Q^2$
compared with our calculations.}
\label{Fig.2}
\end{figure}

\begin{figure}
\centerline{\epsfig{file=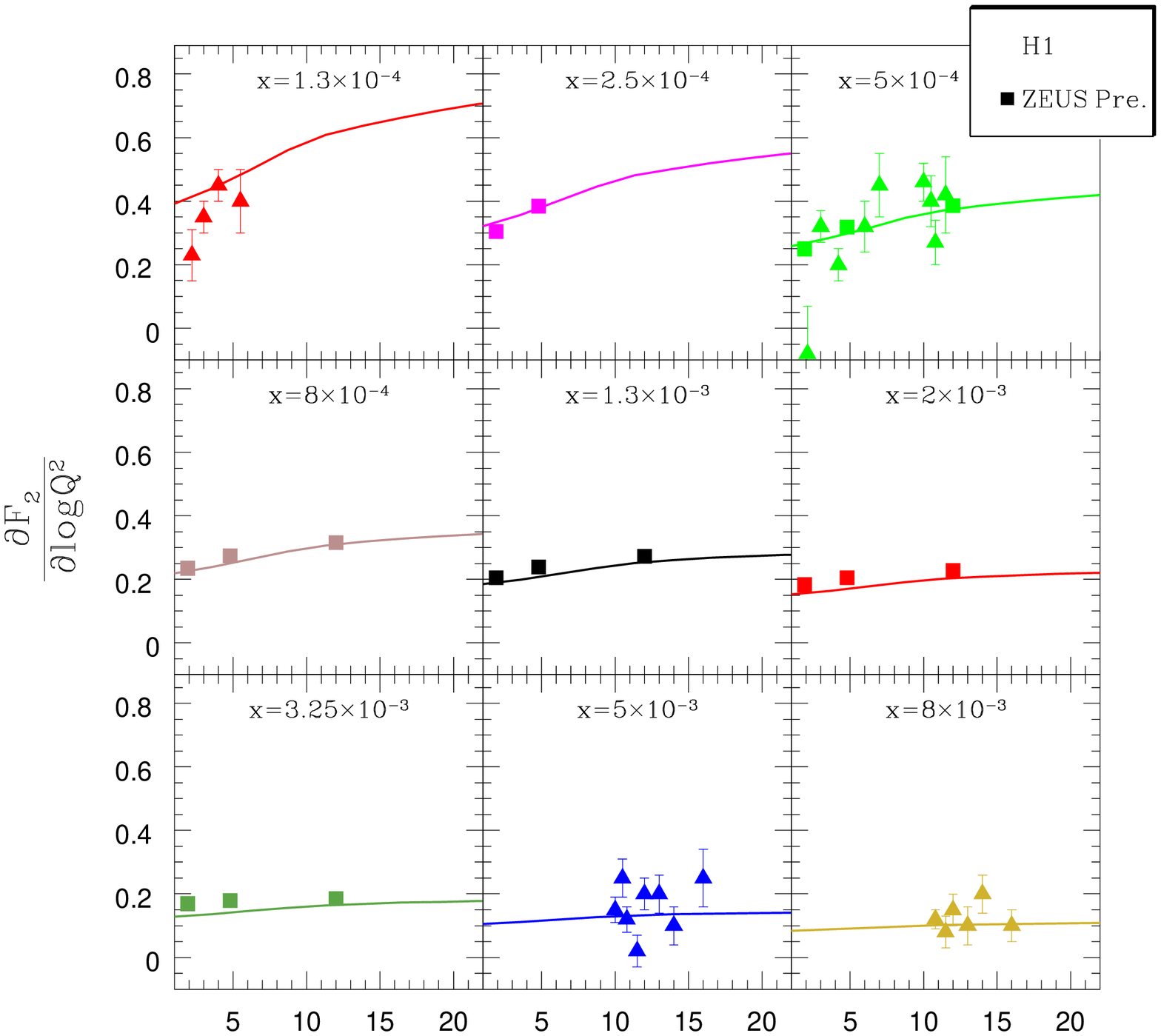,width=170mm}}
\caption{\it $Q^2$ dependence of H1 and ZEUS logarithmic slope data at
fixed $x$ compared with our calculations.}
\label{Fig.3}
\end{figure}

\begin{figure}
\centerline{\epsfig{file=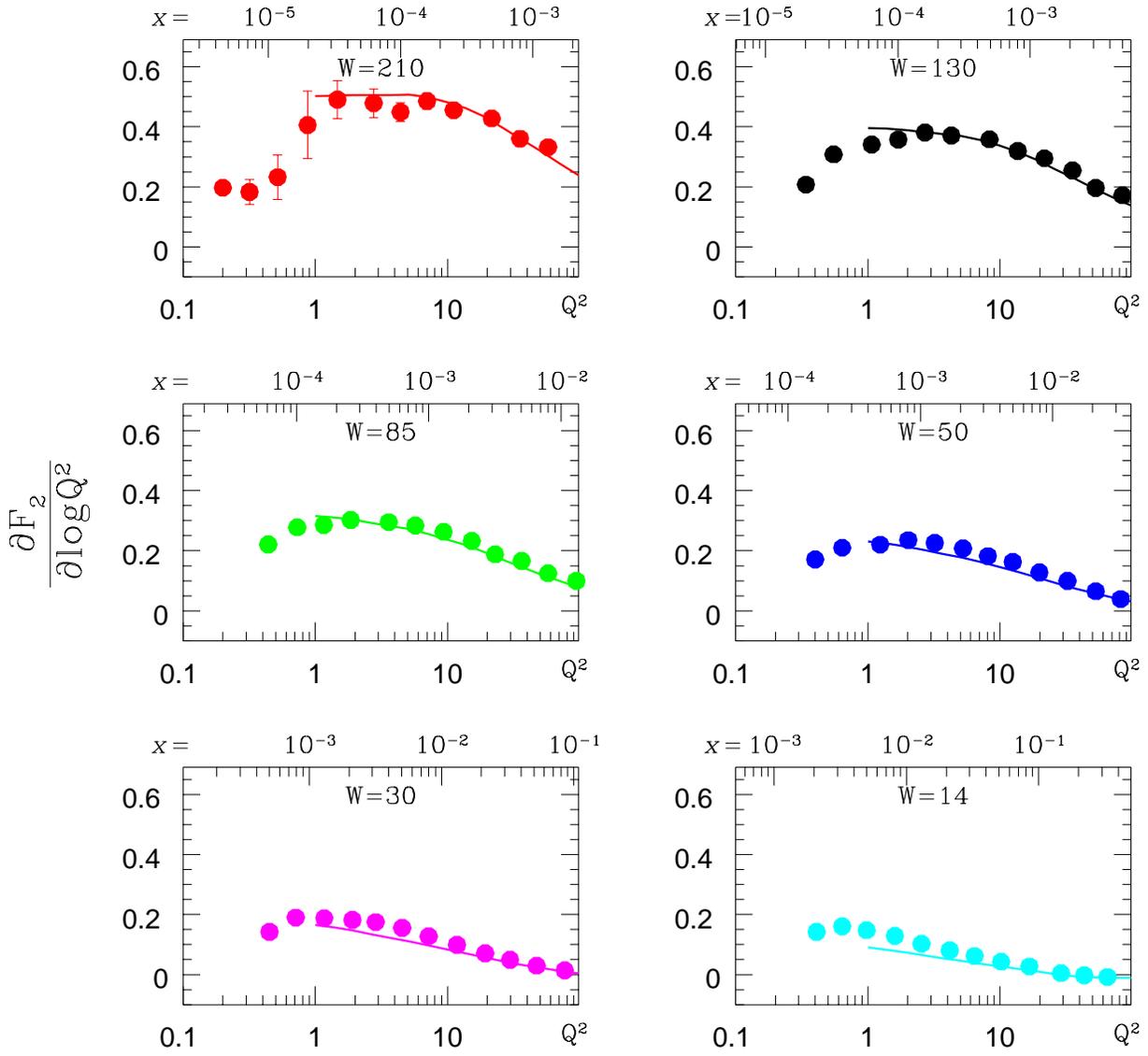,width=170mm}}
\caption{\it ZEUS logarithmic slope data at fixed W compared with our SC
calculation.}
\label{Fig.4}
\end{figure}

\begin{figure}
\centerline{\epsfig{file=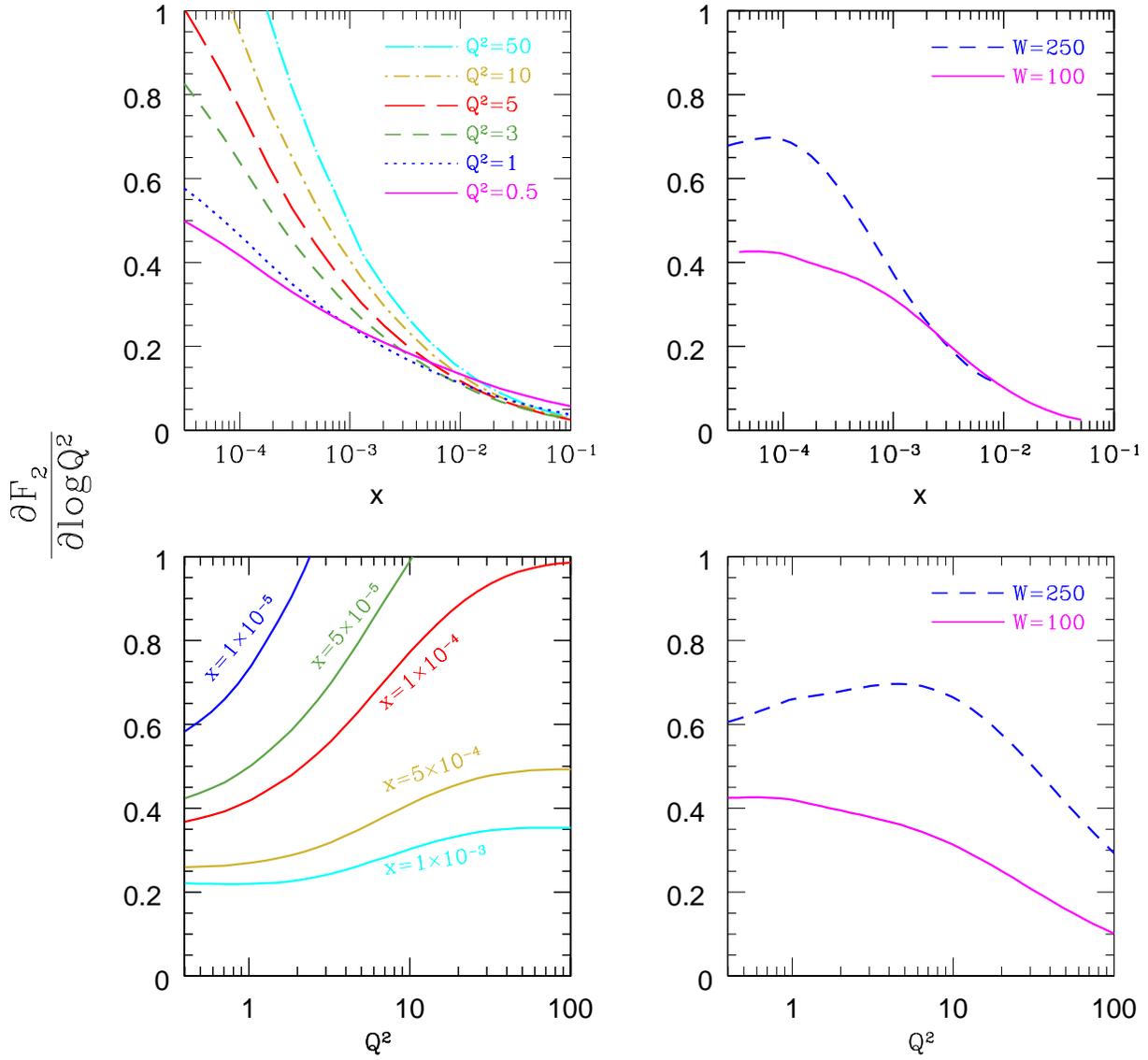,width=170mm}}
\caption{\it Fixed W properties of our SC calculation.}
\label{Fig.5}
\end{figure}

\begin{figure}
\centerline{\epsfig{file=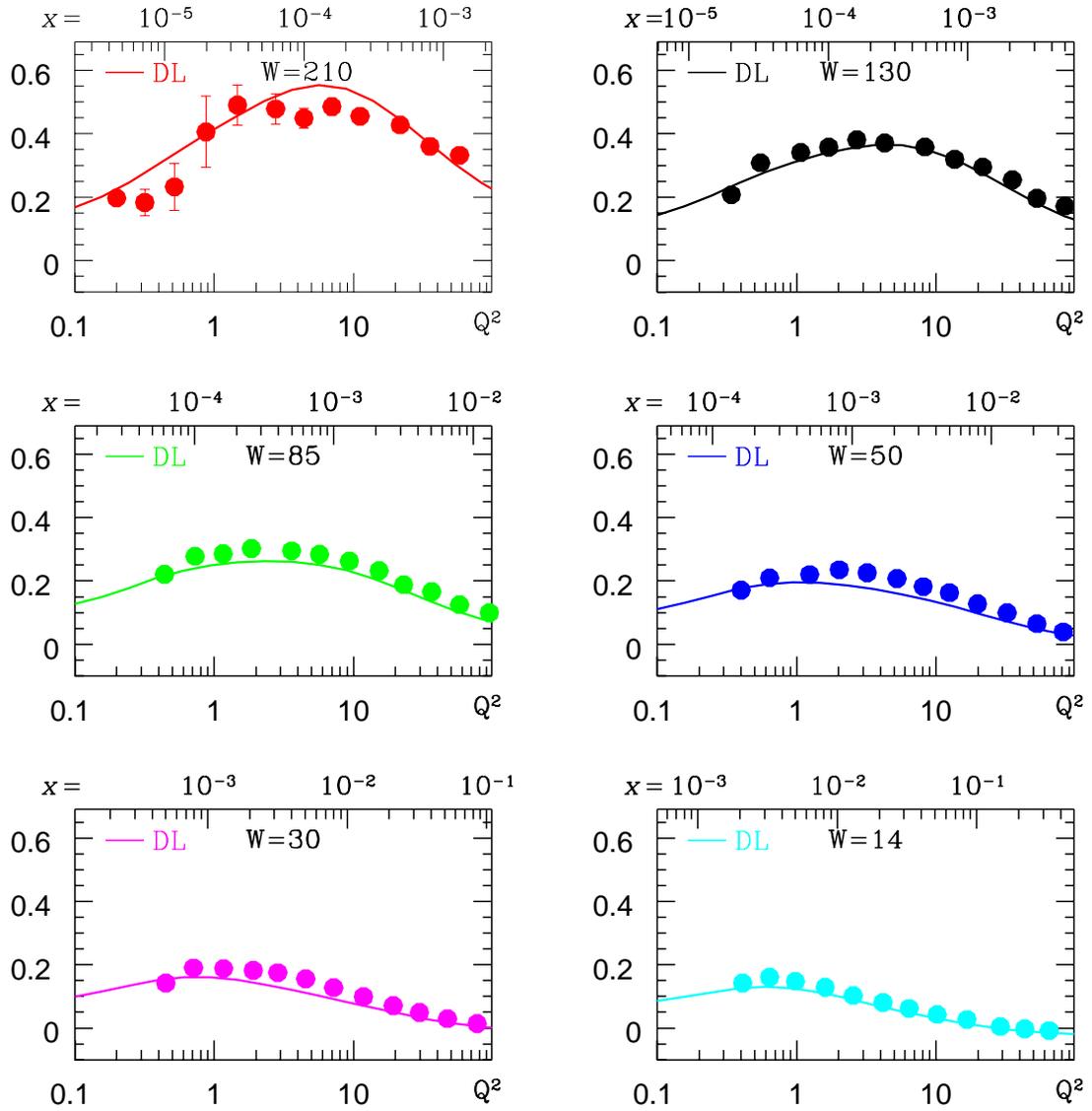,width=170mm}}
\caption{\it ZEUS logarithmic slope data at fixed W compared with DL two
Pomeron model.}
\label{Fig.6}
\end{figure}

\begin{figure}
\centerline{\epsfig{file=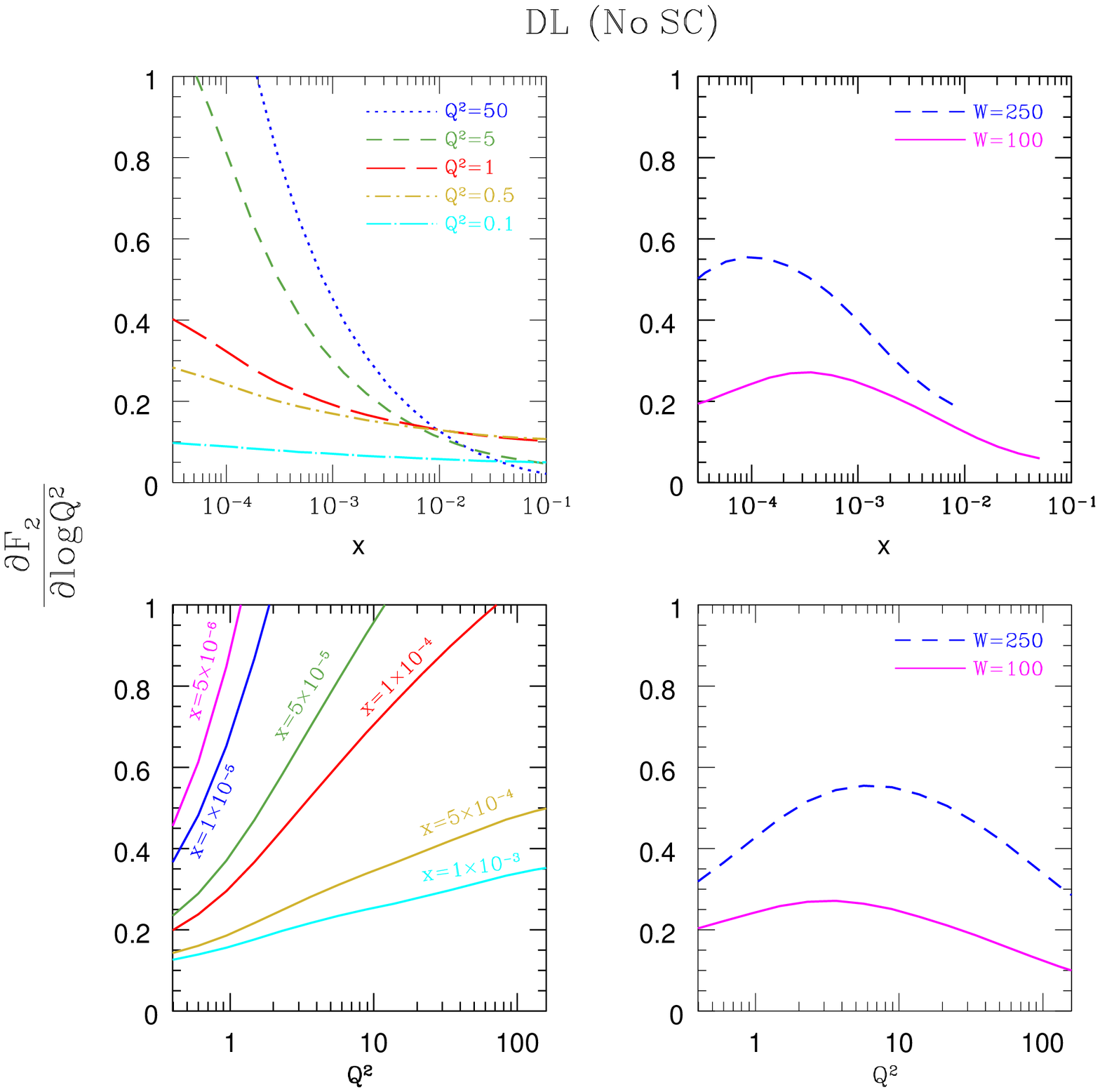,width=170mm}}
\caption{\it Fixed W properties of the DL two Pomeron model.}
\label{Fig.7}
\end{figure}

\section{ Photo and DIS production of $J/\Psi$}

The t=0 differential cross section of photo and DIS production can be
calculated in the dipole LLA approximation 
\cite{RYSKIN}\cite{BRODSKY}\cite{RRML}\cite{FMS}\cite{MRT}.
When using the static non relativistic approximation of the vector
meson wave function \cite{RYSKIN}\cite{RRML}, the differential cross
section has a very simple form, viz.

\beq\label{Q}
\(\frac{d\sigma(\gamma^* p \rightarrow V p)}{dt}\)_0\,=\,
\frac{\pi^3\Gamma_{ee}M_V^3}{48\alpha}\,
\frac{\alpha_S^2({\bar Q}^2}{{\bar Q}^8}\,
\(xG(x,{\bar Q}^2)\)^2 \,\(1\,+\,\frac{Q^2}{M_V^2}\),
\eeq

where in the non relativistic limit we have
\beq\label{Q1}
{\bar Q}^2\,=\,\frac{M_V^2\,+\,Q^2}{4}
\eeq
and
\beq\label{Q2}
x\,=\,\frac{4{\bar Q}^2}{W^2}.
\eeq

%These variables enable us to determine the applicability of pQCD in the
%dipole approximation and thus re-assess the physics presented in the
%previous section.

In the following we discuss in some detail the photo and DIS production of
$J/\Psi$. There is a lot of data available for this channel 
 \cite{ZEUSJ}\cite{H1J}\cite{fixedJ} for the integrated cross section 
spanning  a relatively wide energy  range. From a theoretical point of
view, its hardness (or separation) scale is comparable to the scales we
have studied in our $\partial F_2/\partial ln Q^2$ analysis.  To
relate the integrated cross section to \eq{Q} we need to know $B$, the 
$J/\Psi$ forward differential cross section slope. For this we may use
the experimental values, which are approximately constant. As we shall
see, SC account well for the reported moderate energy 
dependence \cite{ZEUSJ}.
 
The main problem with the theoretical analysis of $J/\Psi$ is the
observation that the simple pQCD calculation needs to be corrected for the
following reasons:
\newline
1) A correction  for the contribution of the real part of the 
production amplitude. 
This correction is well understood and is given by
$C_R^2\,=\,(1\,+\rho^2)$, where 
$\rho\,=\, ReA/ImA\,=\,tg(\frac{\pi \lambda}{2})$ and 
$\lambda\,=\,\partial ln (xG)/\partial ln (\frac{1}{x})$.
\newline
2) A correction  for the contribution of the skewed (off
diagonal) gluon distributions \cite{offdiagonal}. This correction is
calculated to be 
\beq
R_g^2\,=\,
\(\frac{2^{2\lambda+3}\,\Gamma(\lambda+2.5)}{\sqrt{\pi}\,\Gamma(\lambda+4)}\)^2.
\eeq
\newline
3) A more controversial issue relates to the non relativistic
approximation assumed for the $J/\Psi$ Charmonium. Relativistic effects
produced by the Fermi motion of the bound quarks result in a considerable
reduction of the calculated pQCD cross section \cite{Fermi}. We denote
this correction $C_F^2$ and note that it is very sensitive to
the value of $m_c$. Ref.\cite{Fermi} assumes that $m_c \simeq 1.5GeV$ and
obtains $C_F^2\simeq 0.25$ with minimal energy dependence. A small change
in the input value of $m_c$
changes the above estimate significantly. We suggest, therefore, to
consider $C_F^2$ as a free parameter. In our calculations we have used 
$C_F^2=0.66$ which corresponds to a c-quark mass of approximately 
$1.53 GeV$.

Since the $J//\Psi$ photo and DIS cross sections are proportional to 
$\(xG(x,Q^2)\)^2$, the study of this channel can serve as a 
compatibility check supplementing our study of 
$\partial F_2/\partial ln Q^2$. As noted in the introduction our SC
approach was triggered by the observation that none of the latest
p.d.f.'s can provide a good  simultaneous reproduction of the two channels
under consideration.

Our calculation of SC for $J/\Psi$ photo and DIS production is rather
similar to the $Q^2$ logarithmic slope
calculation presented in the previous section. 
We follow our earlier publication \cite{GLMV} and define the damping 
factors due to 
the screening in the quark
sector i.e. the percolation of the $c \bar c$ through the target.
This is given by the following expressions for the longitudinal and
transverse dampings
\beq\label{SCL}
D_{qL}^2\,=\,\frac{\(E_1(\frac{1}{\kappa_q})e^{\frac{1}{\kappa_q}}\)^2}
{\kappa_q^2}
\eeq
and 
\beq\label{SCT}
D_{qT}^2\,=\,
\frac{\(1\,+\,(1\,-\,\frac{1}{\kappa_q})
E_1(\frac{1}{\kappa_q})e^{\frac{1}{\kappa_q}}\)^2}   
{4\kappa_q^2}
\eeq

Our expression for $D_g^2$, the damping in the gluon sector is the square
of the gluon damping defined in the previous section.

Our final expression for the forward cross section is
\beq\label{final}
\(\frac{d\sigma(\gamma^* p \rightarrow J/\Psi p)}{dt}\)_0\,=\,
C_R^2 \cdot C_g^2 \cdot C_F^2 \cdot \(\frac{d\sigma}{dt}\)_0^{pQCD} 
\cdot D_q^2 \cdot D_g^2,
\eeq
where $D_q$ denotes the L and T components as appropriate.

Our calculations as compared with the data are presented in Figs.8 and  9.
 As can be seen, our reproduction of the data is excellent with a
$\chi^2/n.d.f.$ which is well below 1. These excellent $\chi^2$ values are
maintained when calculating over the entire data base as well as limiting
ourselves to the high energy HERA data. Note that $R^2=8.5GeV^{-2}$,
which is the essential parameter in the SC calculation is determined 
directly from the $J/\Psi$ photoproduction forward slope. In a model such
as
ours, we expect a weak dependence of $B_H$ on the energy \cite{GLM2}( see 
Fig.10 ).

\begin{figure}
\centerline{\epsfig{file=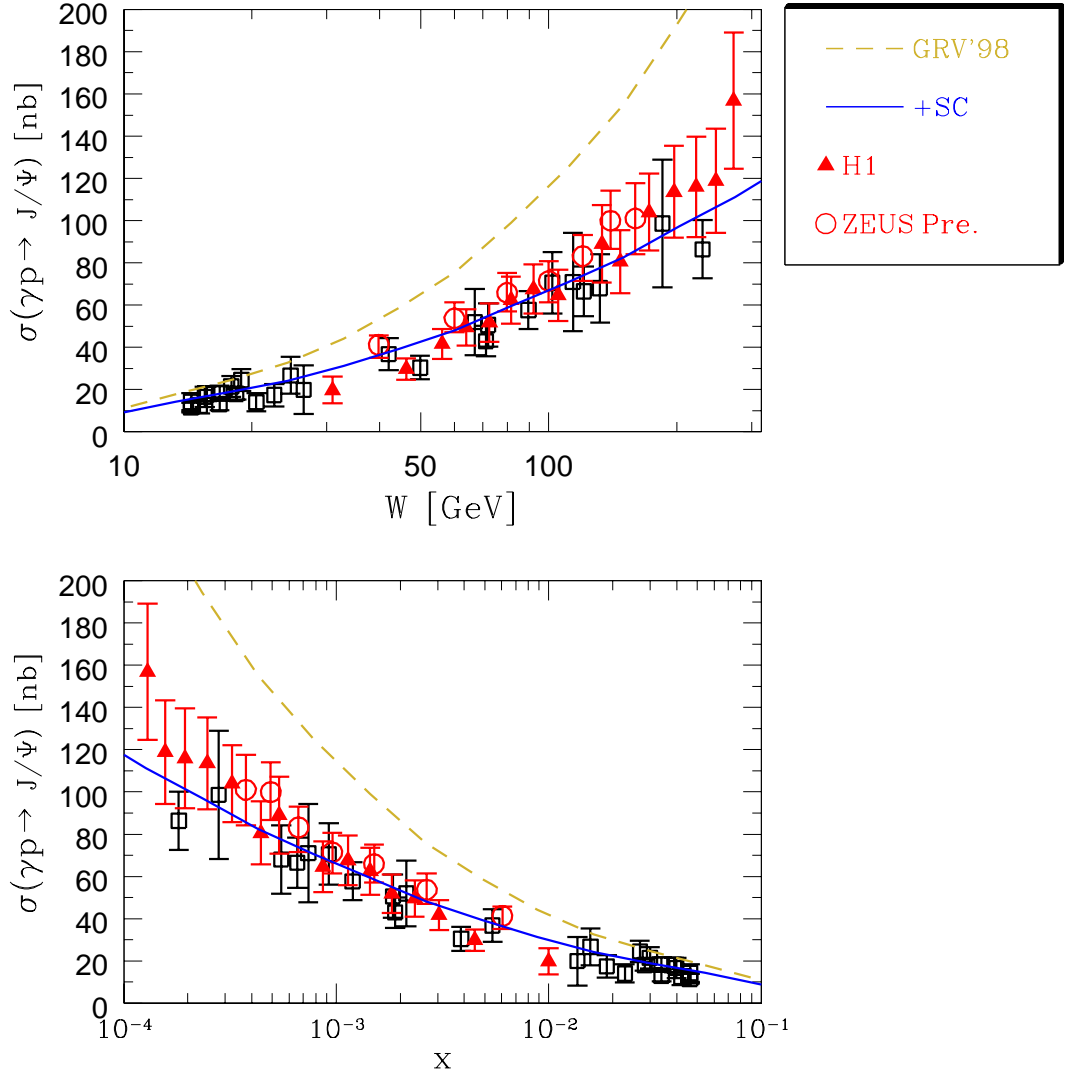,width=170mm}}
\caption{\it Photo production of $J/\Psi$ as a function of W and $x$. Data
and our calculations.}
\label{Fig.8}
\end{figure}

\begin{figure}
\centerline{\epsfig{file=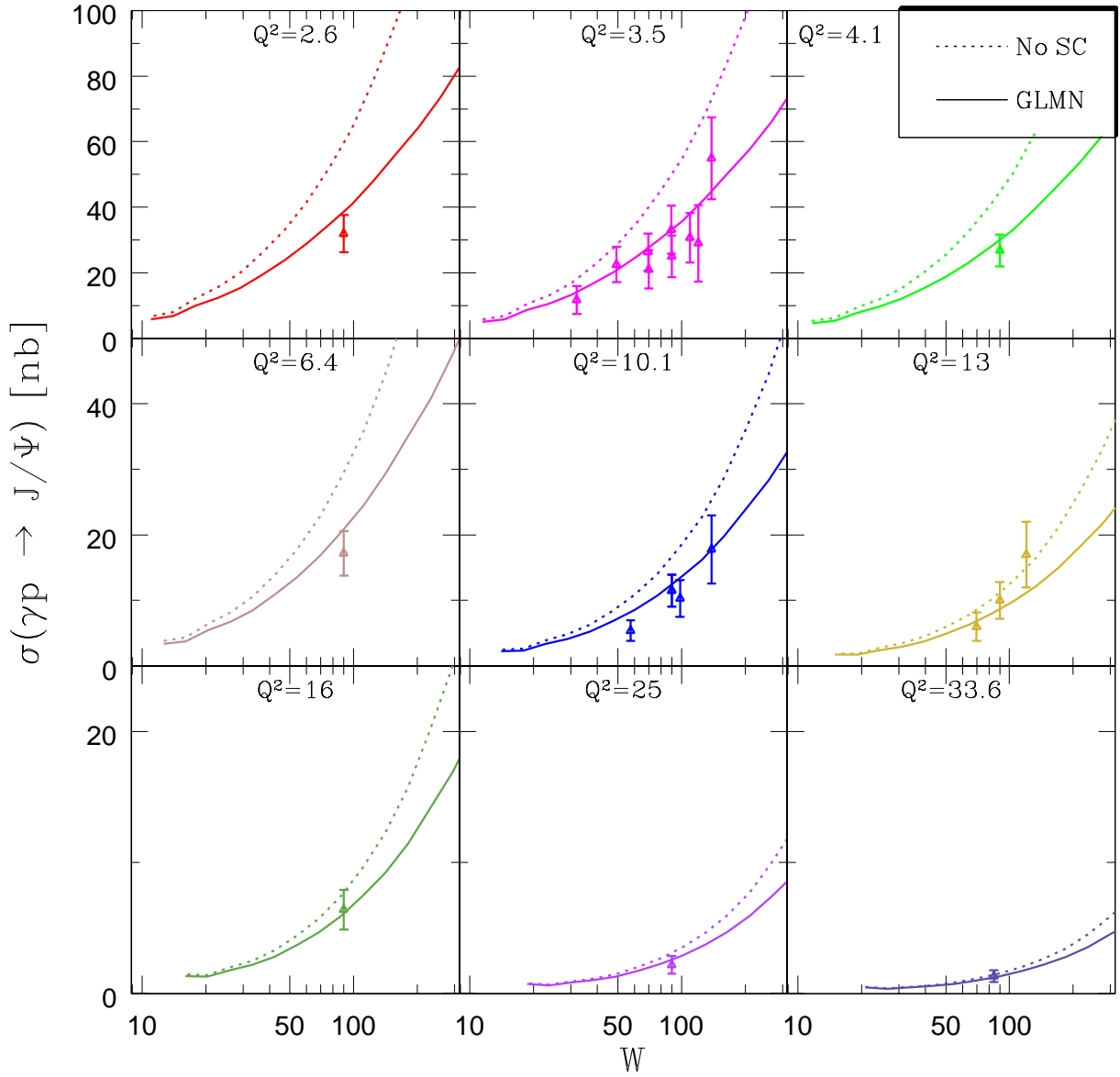,width=170mm}}
\caption{\it DIS production of $J/\Psi$. Data and our calculations.}
\label{Fig.9}
\end{figure}

\begin{figure}
\centerline{\epsfig{file=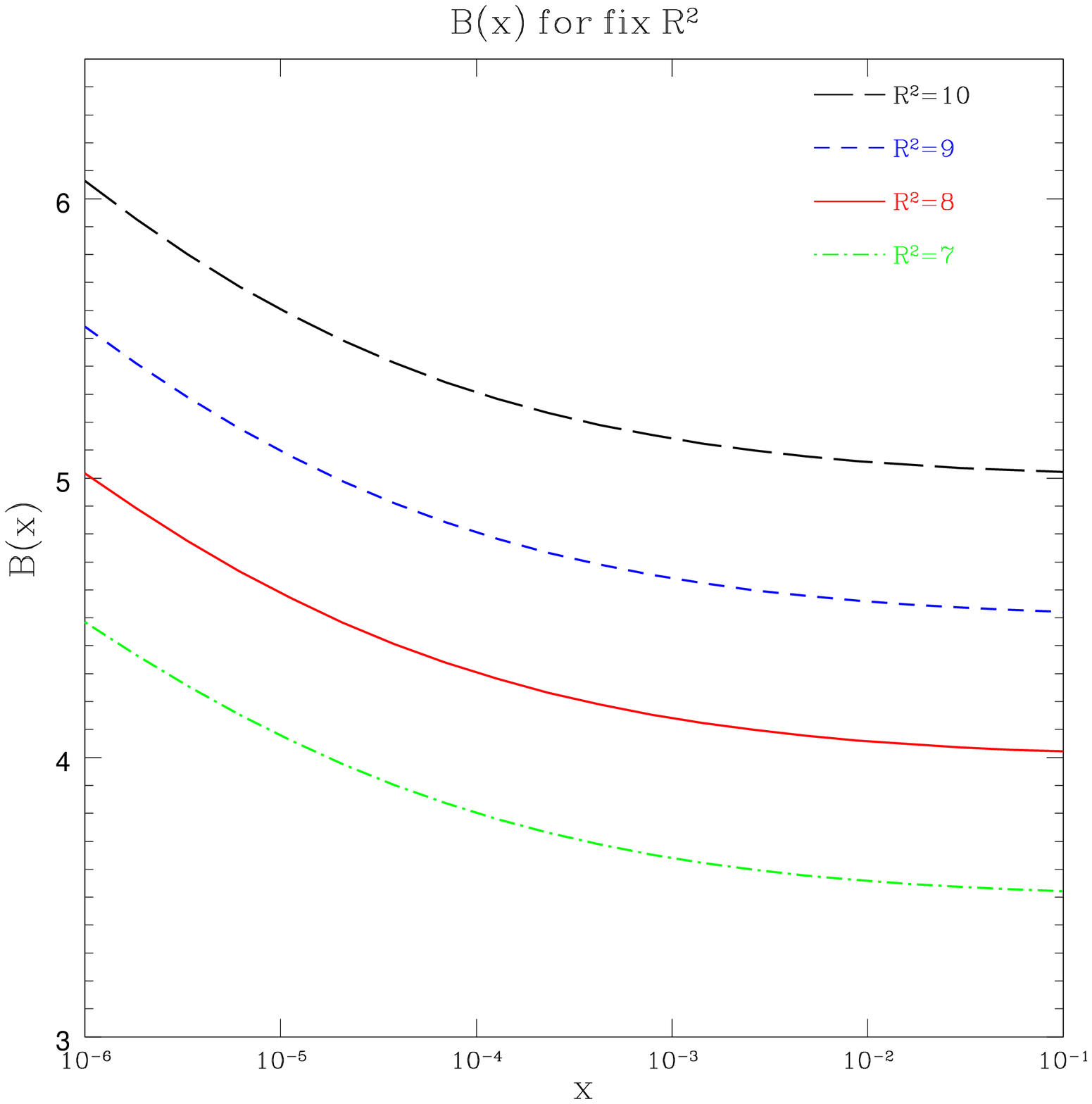,width=170mm}}
\caption{\it The energy dependence of the forward differential slope of
$J/\Psi$ photoproduction.
Data and SC calculations with several values 
of $R^2$.} 
\label{Fig.10} 
\end{figure}

\section{Discussion}

We have shown that NLO GRV98 when screened both  for the percolation of a
$q \bar q$ pair through the target, and for multigluon exchange provides
an excellent repoduction of both the H1 and ZEUS data for the logarithmic
derivative of $F_{2}(x,Q^{2})$ structure function of the proton. The
experimental data and our model are consistent with
$ \partial F_2/ \partial lnQ^2 $ ( at fixed $Q^{2}$ ) being a monotonic 
increasing function of 1/x. No deviation of this behaviour has been seen
even at the lowest values of $Q^{2}$ and x.

The suggested turn over seen in $ \partial F_2/ \partial lnQ^2$ at fixed
W, as a function of x or $Q^{2}$, does not appear to be connected with
saturation effects, and cannot be used as a discriminator, as it appears
in all parametrizations of $F_{2}$ which provide a reasonable description
of the data.

The same screened NLO GRV98 also provides an excellent description    of
the photo and DIS production of $J /\Psi$, once corrections are made for
the real part of the production amplitude, for the skewed ( off diagonal)
gluon distribution and for the Fermi motion of the bound quarks in the
charmonium wave function.

In a seperate publication \cite{GLMNF} we show that we are unable to
obtain  a simultaneous
good description of $\partial F_2/\partial lnQ^2$ and the photo
and DIS production of $J/ \Psi$ using other pdf's on the market
e.g. MRS99,
CTEQ4 and CTEQ5 even after including screening corrections.

\section{Conclusions}
   Only screened NLO GRV98 is able to provide a satisfactory simultaneous
description of the latest HERA data available for
$\partial F_2/ \partial lnQ^2$ and photo and DIS production of $J /\Psi$.

The moral of the paper is that only by a simultaneous analysis of all data
sensitive to shadowing ( saturation ) effects, can one obtain a reliable
estimate of their sizes.

{\bf Acknowledgements:}

UM wishes to thank UFRJ and FAPERJ (Brazil) for their support.

This research was supported by in part by the Israel Academy of Science
and Humanities and by BSF grant \# 98000276.

\end{document}